\documentclass[traditabstract]{aa}
\usepackage{graphicx}
\usepackage{natbib}
\usepackage{txfonts}
\begin{document}

\def\kmsmpc{km~s$^{-1}$~Mpc$^{-1}$}
\def\kms{km~s$^{-1}$}
\def\ho{H$_0$}
\def\h{H1413+117}
\def\wfi{WFI~2033-4723}
\def\sdss{SDSS~J0924+0219}
\def\td{$\Delta$t}

\title{VLT adaptive optics search for  luminous substructures in the lens galaxy towards SDSS~J0924+0219
\thanks{Based on observations obtained with the ESO VLT at Paranal observatory (Prog ID 084.A-0762(A); PI: Meylan). 
Also based in part on observations 
made with the NASA/ESA Hubble Space Telescope, obtained at the Space Telescope 
Science Institute, which is operated by the Association of Universities for Research in Astronomy, 
Inc., under NASA contract NAS 5-26555. These observations are associated with the CASTLES (Cfa-Arizona Space Telescope LEns Survey) survey (ID: 9744, PI: C. S.  Kochanek)}}

\author{ C. Faure\inst{1}, D. Sluse\inst{2,3},  N. Cantale\inst{1},  M. Tewes\inst{1}, F. Courbin\inst{1},  
P. Durrer\inst{1} and  G. Meylan\inst{1} }

\institute{Laboratoire d'astrophysique, Ecole Polytechnique F\'ed\'erale de Lausanne (EPFL), Observatoire de Sauverny, 1290 Versoix, Switzerland
\and
Astronomisches Rechen-Institut am Zentrum f\"ur Astronomie der Universit\"at Heidelberg, M\"onchhofstrasse  
12-14, 69120 Heidelberg, Germany
\and
Argelander-Institut  f\"ur  Astronomie, Auf dem H\"ugel 71, 53121 Bonn, Germany
}
\date{Submitted: 01/08/2011, Accepted: 29/09/2011}

\authorrunning{Faure et al.}
\titlerunning{A search for luminous substructures towards \sdss}
\abstract{Anomalous flux ratios between quasar images are suspected to be caused by substructures in lens galaxies. We present new deep and  high resolution H and Ks imaging of the strongly lensed quasar \sdss\ 
obtained using the ESO VLT with adaptive optics and the Laser Guide Star system. \sdss\ is particularly interesting as the observed flux ratio between the quasar images vastly disagree with the 
predictions from smooth mass models.  With 
our adaptive optics observations we find a luminous object, Object L, located $\sim 0.3$\arcsec\ to the North of the lens galaxy,
but we show that it can not be responsible for the anomalous flux ratios. Object L  as well as a luminous extension of the lens galaxy to the South are  seen in the archival HST/ACS
image in the F814W filter. This  suggests that  Object L is part of a bar in the lens galaxy, as also 
supported by the presence of a significant disk component in the light profile of the lens galaxy. Finally, we do not find
evidence for any other luminous substructure that may explain the quasar images flux ratios. However, owe to the persistence of the flux ratio anomaly over time ($\sim $ 7 years) a combination of microlensing  and milli-lensing is the favorite explanation for the observations.}

\keywords{Gravitational lensing: strong, Galaxies: quasars:individual:SDSS~J092455.87+021924.9}
\maketitle

\section{Introduction}

Strong gravitational lensing is now part of the standard toolbox of the astrophysicist. 
It consists in a powerful test for cosmological models 
\citep[e.g.][]{barnabe2011, suyu2010a, coemoustakas2009, fedeli2008, fedelibartelmann2007} 
and allows us to study the distribution of luminous and dark matter in galaxies as well as the
evolution of their properties with redshift  \citep[e.g.][]{ruff2011, faure2011, tortora2010, Auger2010,koopmans2006}. 

Such studies require a model for the total projected potential well of  lensing galaxies, which is often described as a 
smooth 2D elliptical profile, with or without contribution of intervening objects along the line of sight. 
However, the increasing accuracy of the observational constraints provided by high resolution 
imaging, has quickly shown the limitations of such simple models: in quadruply imaged quasars,
smooth models are often unable to account simultaneously for the milli-arcsec astrometry  and for the near- and mid-IR flux ratios of the
quasar images
\citep{Vegetti2010b, chantry2010, yoo2005, biggs2004, kochanekdalal2004, keeton2003, metcalfzhao2002,koopmans2002, mao1998}.

The reason for the observed discrepancies between the model predictions and the measurements
may be  the presence of substructures in the halo of lensing galaxies. Such small deviations to the smooth 
potential might affect significantly the predicted flux ratios while barely changing the astrometry of the quasar images. 
This is especially true when a quasar image is a saddle point of the arrival time surface \citep{Schech2002}. 
Strong lensing therefore offers a sensitive way
to indirectly detect  and weight substructures in galaxy halos, whether luminous or not \citep[e.g.][]{vegetti2010a,Suyu2010b, mckean2007}.

In the present work, we have searched  for substructures in the halo of the lens galaxy towards the quadruply 
imaged quasar with the most anomalous flux ratios, \sdss\  \citep{inada2003}, at z$_{\rm quasar}=1.524$, discovered 
in the Sloan Digital Sky Survey \cite[SDSS,][]{york2000}.  Hubble Space Telescope (HST) spectra of the system 
reveal that one of the quasar images, labelled D in  \cite{inada2003} (see also Fig.~\ref{step5}), is extremely faint \citep{keeton2006}. It is also a saddle point in the arrival time surface, hence making it more likely to be demagnified owe to 
micro- and milli-lensing \citep{Schech2002}.   The microlensing hypothesis provides a satisfying explanation for the observed flux ratio anomaly \citep{bate2011, mediavilla2009, morgan2006, keeton2006}. However, this flux anomaly lasts for at least 7 years, which is significantly more than the    expected duration of microlensing fluctuations in this lens (0.39 years), and larger than the duration averaged over a sample of 87 lenses, i.e, 7.3 month,  \citep{mosquera2011}. It is therefore conceivable that not only stellar-microlensing takes place in this system but also that a more extended/massive substructure also contributes to the observed anomaly.
A deep spectrum of the lens galaxy was obtained with the ESO Very Large Telescope (VLT) 
and the FORS1 instrument \citep{eigenbrod2006}, leading to  z$_{\rm lens}=0.394\pm0.001$ from stellar absorption lines.  In addition \citet{eigenbrod2006} notice numerous elongated features within and around the Einstein ring in the HST images. This observation leads them to conclude that the lensed source may be double.

In order to investigate further the nature of these features and to look for possible faint and small satellites to the lensing
galaxy, we conduct deep Adaptive Optics (AO) imaging 
observations of \sdss\ with the VLT.  The paper is organized as follow: in Section~\ref{obs}, we present the  VLT AO observations
and the  HST dataset.  In  Section~\ref{search} we detail our findings of a luminous substructure in the AO images.
In  Section~\ref{acs}, we show that the substructure is also present in the optical, from HST/ACS observations. 
In  Section~\ref {nat}, we  discuss the possible nature of the substructure.  Finally, we present our 
conclusions in Section~\ref{conc}.

Throughout this paper, the WMAP5 $\Lambda$CDM cosmology  is assumed ($\Omega_{\rm m}=0.258$, 
$\Omega_\Lambda=0.742$, $H_0=72$ \kmsmpc). All magnitudes are in the AB system.

\section{The datasets}\label{obs}

\subsection{Adaptive optics imaging with the VLT and NACO}\label{nana}

We observed \sdss\ with the near-infrared ca\-mera CONICA, which is mounted on the Adaptive Optics 
system NAOS, NACO for short, installed at the Nasmyth B focus of the VLT-UT4 at the ESO Paranal Observatory, Chile. Our observations 
were obtained on 2010 March 12,  and 14 (H- and  Ks-bands) and on 2009 December 14 (Ks-band), using the laser guide star 
facility. The star U0900-06393922 ($V=16.5$mag), located 47\arcsec\ away from our target, was used to correct for  ``tip-tilt". 
The S27 CONICA camera has a pixel size of $x=27.053\pm0.019$~mas  and a field-of-view of 28\arcsec\ on a side.
A total of 30 (41) exposures of $3 \times 60$s  (NDIT$\times$DIT)  was obtained in the Ks- (H-) band, with airmasses in the range 
$\sec(z)=1.1-1.3$. The weather conditions were photometric.

The reduction procedure follows the steps exposed in \cite{sluse2008}, i.e., we first subtract dark frames to all the
near-IR science images and we apply standard flat-fielding. We use twilight flat-fields obtained at most 4 days before 
the observations. We check for the temporal stability of these calibrations: flat-fields obtained 4 days apart are 
undistinguishable while flat-fields obtained 4 months apart show differences $<$ 2\%.

The frames are divided in subsets of continuous observations. Each frame is sky-subtracted using the the 
{\texttt{xdimsum}} IRAF{\footnote
{IRAF is distributed by the National Optical Astronomy Observatories, which are operated by the Association of Universities for 
Research in Astronomy, Inc., under cooperative agreement with the National Science Foundation.}} package. Each image is visually
inspected to exclude data with unreliable sky subtraction, with residual cosmic-rays, or with too heavily distorted 
Point Spread Function (PSF). Finally, we construct two deep images by combining 27 (28) frames in the Ks-band 
(H-band) using the {\texttt{xnregistar}} task. The total equivalent exposure time  is 4860 s in the Ks filter and 5040~s in the 
H filter.  

The PSFs of the combined images has a Full Width at Half Maximum of ${\rm FWHM}\sim 0.25$\arcsec\ in the H-band and 
${\rm FWHM}\sim 0.20$\arcsec\ in  the Ks-band, with a moderate ellipticity, $\epsilon\sim0.1$. Based on the observations of
standard stars on 2009 December 13 and on 2010 March 13, we derive photometric zeropoints: 
${\rm ZP(Ks)}=23.05 \pm0.15$~mag s$^{-1}$ and ${\rm ZP(H)}=24.06\pm0.10$~mag s$^{-1}$. 
The large uncertainties on the zeropoints reflect mainly the temporal variations of the low frequency wings of the 
PSF and angular anisoplanatism \citep[see e.g.][]{esslingeredmunds1998}. 
Extinction corrections of 0.034~mag in the H-band and 0.043~mag in the Ks-band  are 
used to flux-calibrate the science frames \citep[][]{lombardi2011}.

\subsection{HST optical imaging with the ACS}
 
Optical HST images of \sdss\ are available from the archives (ID: 9744, PI: C. S.  Kochanek). These data consist of F814W and F555W drizzled 
images obtained with the Wide Field aperture of the Advanced Camera for Surveys (ACS).
The four F814W drizzled images were initially taken on 2003  November 18 and 19 with an exposure time of 
574~s each. The four F555W frames were  observed in 2003  November 18 with a  total exposure time of 547~s each.  
The images have been only recently drizzled  \citep[April 2010, ][]{koekemoer2002}. The non-drizzled dataset was presented for the first time 
by \cite{keeton2006} and used in \cite{eigenbrod2006}. In the following, we use the latest drizzled images directly from the archives. The pixel size in
these images is 0.05\arcsec.

\begin{figure}[t!]
\begin{center}
\includegraphics[width=9cm]{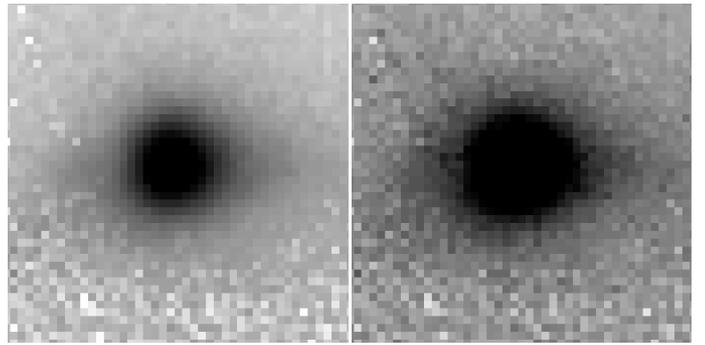}
\caption{ PSFs created for the NACO Ks-band image. {\it Left:} zero order estimate of the PSF, i.e., 
PSF$_0$ (see text). {\it Right:} PSF$_5$, displayed here with the same grey scale as PSF$_0$ (logarithmic scale).}
\label{figpsf}
\end{center}
\end{figure}

\begin{figure}[t!]
\begin{center}
\includegraphics[width=9cm]{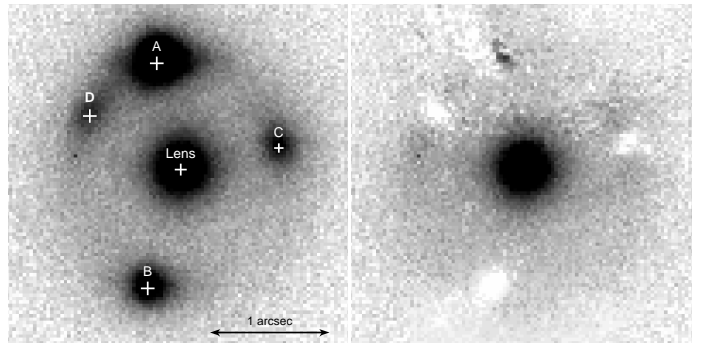}
\caption{NACO Ks-band image of \sdss.  {\it Left:} combined frame, where the quasar images are labelled following 
\citet{inada2003}. {\it Right:} same image where the quasar images have been fitted using PSF$_5$ and subtracted 
using {\tt Galfit}. The residuals at the quasar positions are negligible.}
\label{step5}
\end{center}
\end{figure}

\section{Analysis of the NACO images}
\label{search}

Our goal with the AO images is to detect any possible faint and small satellite
to the lens galaxy. Given the spatial scales involved, such satellites are hidden in the glare of
the quasar images or of the lens galaxy itself. It is therefore mandatory to estimate properly the 
AO PSF and to subtract the light contribution of the quasar images and of the lens galaxy from the data.

\subsection{PSF construction}
\label{secpsf}

With the small field of view of the NACO instrument, no bright star is available to measure the PSF. 
We therefore create a PSF using the quasar images themselves. For that purpose, we use the following iterative method.

First, we estimate a guess PSF by combining the images of the 3 brightest quasar images. Lets call it PSF$_0$. The 3 quasar images are aligned and normalized to their peak intensity. PSF$_0$ is built by assigning, at every pixel location, the lowest pixel value among the 3 quasar images. 
PSF$_0$ is then used in {\tt Galfit} \citep{peng2010,peng2002} to fit  and 
subtract the 4 quasar images. The resulting image obviously shows strong residuals at each quasar position.
We median-combine these residuals,  previously normalized and aligned. The result of this operation is then weighted by the square root of PSF$_0$, and added to the latter to build PSF$_1$, our next estimate of the PSF.
This relatively arbitrary weighting allows us to minimize the noise contamination of the pixels located far away from the PSF center, i.e., where PSF$_0$ is already a good approximation of the true PSF.

 
Second,  we fit the galaxy light profile using a Sersic  plus an exponential disk,  in addition to the 4 quasar images. 
This light model is convolved by PSF$_1$ during the fit. 
The fit of the lens galaxy is not possible in the first step of the PSF creation because the poor quality of PSF$_0$ 
prevents {\tt Galfit} to converge properly. We note that the use of an exponential disk in addition to the Sersic 
profile to model the lens galaxy is  crucial to obtain acceptable fits. The residual image obtained with the improved 
PSF and lens model is used to compute a new  estimate of the PSF, that we call PSF$_2$.

We then repeat this second step until the residuals to the fit at the quasar positions can not be improved anymore. 
A typical number of 5 cycles is necessary to achieve this goal. We show in Fig.~\ref{figpsf} the PSF obtained after 
5 iterations, for the Ks-band image. As mentioned  in Section~\ref{nana}, the  PSF ellipticity is $\epsilon\sim0.1$. The apparent elongation of PSF$_5$  in Fig.~\ref{figpsf} is noticeable only at large radius and is visible because of the logarithmic scale. It is also typical of the NACO PSF shape, as  observed in other targets of our observational program. 
Fig.~\ref{step5}, shows the Ks-band image after subtraction of the quasar images using
PSF$_5$. We do not observe systematic residuals under the quasar images, hence suggesting that  PSF$_5$ is reliable. We do not detect any luminous object under the quasar wings. 

Note that the 4 quasar images are located at very different positions on the Einstein ring. The quasar host
galaxy has therefore very different Position Angles (PA) at the four quasar positions. This fortunate configuration has
the consequence to minimize the quasar host galaxy contribution when median-averaging the residuals at each step of the
PSF construction. 
 
\subsection{Quasar image fits }
The quasar image fits include light from the ring (Fig.~\ref{step5}). This prevents us from measuring accurate photometry for the quasar images. Rough estimates of the magnitude differences between images A and D are $\Delta({\rm M}_{\rm D}-{\rm M}_{\rm A})_{\rm H}\sim3.3$~mag and $\Delta({\rm M}_{\rm D}-{\rm M}_{\rm A})_{\rm K_s}\sim2.2$~mag. They confirm the flux anomaly observed by \cite{inada2003}, even at longer wavelength.

\subsection {Galaxy light profile fit and subtraction}\label{secgalf}

\begin{figure*}[htbp]
\begin{center} 
\includegraphics[width=18cm]{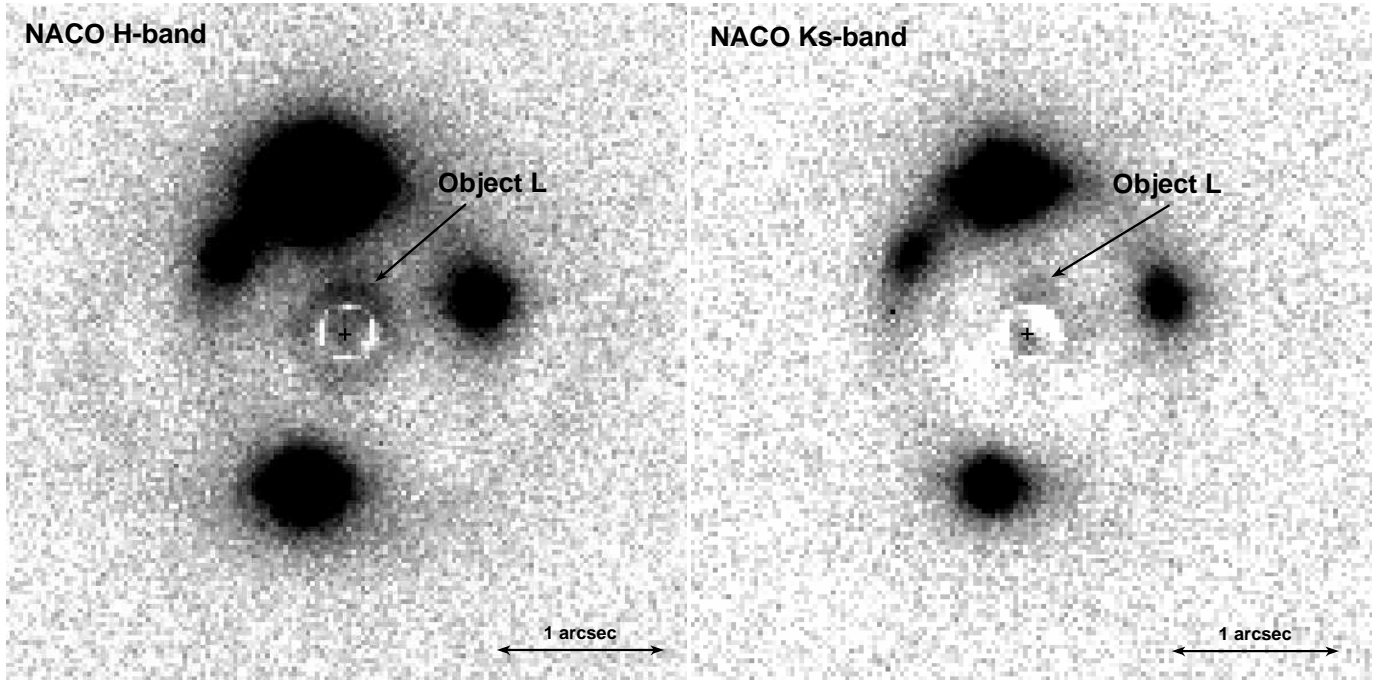}
\caption{VLT/NACO H- and Ks-band images of \sdss\ after subtraction of the lens galaxy with {\tt Galfit}.  
North is to the top, East is to the left. 
The crosses indicate the fitted position of the lens.  
In both bands we identify  a luminous residual not fitted by the Sersic profiles (Object L)  located about 
0.27\arcsec\ to the North of the lens center. }
\label{figfit}
\end{center}
\end{figure*}

We use {\tt Galfit} to fit  and subtract the lens galaxy from the data. We are primarily interested in a clean subtraction of the lens galaxy light within the Einstein radius. To do so, we 
mask the area of  the Einstein ring  (quasars and quasar host galaxy). The use of masks ensure that {\tt Galfit} does not try to fit the ring as a component of the lens galaxy. However, it can be a source of inaccuracies, for example if the zone masked encompasses the galaxy effective radius.  
 
The PSFs obtained in Section~\ref{secpsf}  are used to convolve the analytic profiles in {\tt Galfit} in the
H and Ks bands. The best fits are obtained when the lens galaxy is described as 
the sum of two Sersic profiles (see Fig.~\ref{figfit}). The
first one has a very shallow index, ${\rm n}_1=0.13\pm0.04$ and an effective radius ${\rm R}_{{\rm eff},1} =0.90\pm0.01$\arcsec. The
second profile is much steeper, with ${\rm n}_2=7.17\pm0.65$ and has a smaller effective radius, 
${\rm R}_{{\rm eff},2}=0.35\pm0.04$\arcsec.   
In the fits, the two profiles are separated by $\delta_{1,2}=30\pm8$~mas and have ${\rm PA}_{1,2}\sim -80^\circ$ 
(positive North to East). Their ellipticity is very small, $\epsilon_{1,2}\lesssim 0.1$. The  values given above 
are averaged over the two near-IR bands and the error bars  correspond to the dispersion between those values. The agreement between the fits in the two band is remarkable.  The total magnitude of the lens galaxy  is ${\rm Ks}=15.75\pm0.15$~mag and ${\rm H}=16.24\pm0.10$~mag. 
 
According to our image decomposition with {\tt Galfit}  the lens galaxy contains a significant disk-like structure,
as suggested by the necessity to include a shallow Sersic profile in the fit. This is also suggested by earlier results 
in 3 HST bands \citep[][]{eigenbrod2006}. We will discuss further this finding in Section~\ref{nat}.  
After subtracting our model of the lens galaxy (Fig.~\ref{figfit}), we detect a faint but significant object North to the 
galaxy's centroid. The detection is significant both in the H and Ks filters : ${\rm Ks}=22.20\pm0.15$~mag, ${\rm H}=23.23\pm0.10$~mag, which translates in ${\rm L}_{\rm V}\sim4\pm1\times10^6$~${\rm L}_{\odot,{\rm V}}$. This luminosity is K-corrected  assuming an elliptical spectral energy distribution. We also corrected for extinction and evolution. In the following, we refer to this possible
substructure of the lens as  Object L.

\begin{table}[t!]
\caption{Relative position of the quasar images, of the lens galaxy  and of Object L to image A, as measured in the ACS/F814W deconvolved images. The quasar images are labelled as in Fig.~\ref{step5}.   The astrometric uncertainties for the quasar images reflect the internal error of the deconvolution method. For the lens galaxy they represent the dispersion in the Sersic centroids for different acceptable fits. } \label{pos}
\begin{center}
\begin{tabular}{c| c c c  }
\hline
& $\Delta$RA & $\Delta$DEC \\
           & {\it \arcsec}&{\it \arcsec}     \\
\hline
\hline
A & $0.0\pm 0.001$ & $0.0\pm 0.001$ \\
B & $+0.068\pm 0.001$ & $-1.804\pm 0.001$\\
C & $-0.957\pm 0.001$ & $-0.701 \pm 0.001$ \\
D & $+0.555\pm 0.006$ & $-0.422\pm 0.006$ \\
Lens &$-0.176\pm  0.020$   & $-0.860\pm 0.020$\\
Obj. L& $-0.15\pm 0.02$  &  $-0.52\pm 0.01$\\
 \hline                         
\end{tabular}
\end{center}
\label{default}
\end{table}%

\section {Analysis of the HST images}
\label{acs}

\begin{figure*}[htbp]
\begin{center}
\includegraphics[width=18cm]{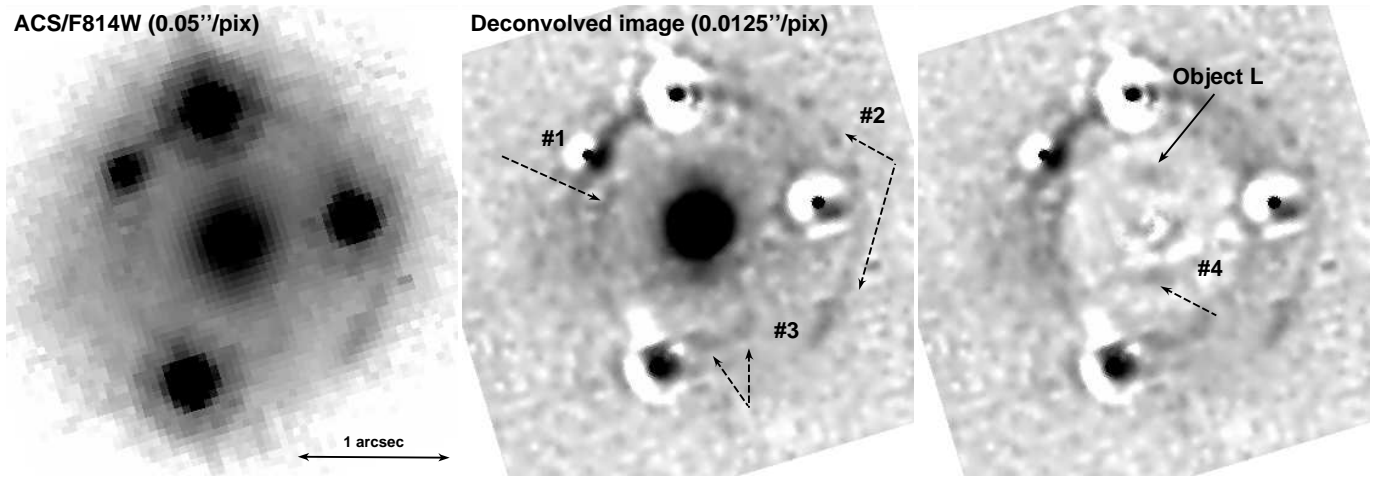}
\caption{HST/ACS image of \sdss\ in the F814W filter. {\it Left:} original drizzled image obtained from the archive. {\it Middle:}  deconvolved image. The pixel size is 4 times smaller than in the left panel. {\it Right:} deconvolved image, where the lensing 
galaxy profile has been fitted and removed using {\tt Galfit}.  The arc-like features \#1-4 are discussed in the text. North is to the top, 
East is to the left.}
\label{figfitacs}
\end{center}
\end{figure*}

The luminous Object L discovered in the NACO images is also visi\-ble in the ACS/F814W drizzled images. 
As the field of view of the ACS is much wider than that of NACO several stars are available to build a reliable PSF 
and to spatially deconvolve the images. To do so, we use the MCS algorithm \citep{magain1998}. 
The required PSF  is built using   3 stars located within a 2\arcmin\ radius from the lens. The spatial resolution 
after deconvolution is FWHM=0.025\arcsec. The adopted pixel size in the deconvolved data is 
a fourth of the original data, i.e,  0.0125\arcsec.   The original F814W drizzled image and its deconvolved version
are displayed in Fig.~\ref{figfitacs}. We do not consider here the much shallower F555W images. 

We then fit the deconvolved  light profile of the lens galaxy, using a mask at the position of the ring and of the quasar images. As was the case with the VLT/NACO data, 
two Sersic profiles  are necessary to obtain good results. 
The parameters of the best fit Sersics are :  ${\rm n}_1= 0.06$, ${\rm R}_{{\rm eff},1}=0.43$\arcsec, ${\rm PA}_{1}=9^\circ$ for the shallow profile, 
and  ${\rm n}_2=0.58$, ${\rm R}_{{\rm eff},2}=0.14$\arcsec, ${\rm PA}_{2}=-20^\circ$ for the steeper profile. Their ellipticity is low  
($\epsilon_{1,2}\lesssim 0.1$) and their centroid are separated  by  $\delta_{1,2}=4$~mas.
The total luminosity of the lens galaxy is ${\rm F814W}=19.36\pm0.01$~mag.
The profiles are very different from those fitted in the NACO images. This might owe to the different depths, wavelength
ranges and stellar populations probed by the data,  but also to the different masks used during the fits and to the fact that the HST image (contrary to the NACO image), was preliminarily deconvolved before the use of {\tt Galfit}. As a consequence the photometry of the lens galaxy is possibly affected by systematics errors of several tenth of a magnitude. This result  illustrates well the difficulty to measure the light profile of a lens galaxy. However, we notice that both the HST and  NACO data require a disk-like component 
in the fit.
In addition to Object L, clearly visible in the HST optical image, we show four arc-like features in Fig.~\ref{figfitacs}. 
Features \#1, \#2 and \#3 have been discovered earlier in the non-drizzled image \citep{eigenbrod2006}. They were 
successfully modeled as a secondary set of images at the quasar redshift. Feature \#4 is identified in the ACS/F814W 
drizzled image for the first time.  The astrometric measurements for the quasar images, the lens galaxy and Object L are given in Table~\ref{pos}. We measure ${\rm F814W}= 25.50\pm0.02$~mag for Object L, which translates into a rest frame luminosity  ${\rm L}_{\rm V}\sim9.4\times10^7$~${\rm L}_{\odot,{\rm V}}$ (see also Section~\ref{secgalf}).

Finally, we note that the orientation of the HST  image is about $-70^\circ$. Therefore, Object L  can not  be a spike 
produced, e.g., by quasar image A or by the sharp nucleus of the lens galaxy. Removal of these spikes is in fact the main
motivation for carrying out image deconvolution. In addition,  Object L is present both  in the space image and in the 
NACO images, i.e., in 3 independent bands with two very different instruments,  making it unlikely to be an artifact in 
all three datasets. 
    
\section{The nature of the luminous substructure}
\label{nat}

The faint Object L seen in the VLT and HST images may be interpreted in several ways. It can be 
(i) a satellite of the lens galaxy,  (ii) the fifth central 
image of the lensed quasar, (iii)  the lensed image of a source unrelated to the quasar, or (iv)
the brightest part of a bar in the lens galaxy.
In the following, we test the different hypothesis  using simple lens modeling.


  
(i) Our model consists in a smooth analytical mass profile to which we add a small perturber using 
the {\tt Lenstool} software \citep{kneib1993,jullo2007}. The main component of the model is a Singular 
Isothermal Ellipsoid (SIE) with an additional contribution of an external shear.  The parameters to 
be fitted are the SIE position, its velocity dispersion, and the external shear strength and direction. 
The observational constraints are  the positions of the quasar images available from the HST images (Table~\ref{pos}). In addition, we use the ellipticity measured as a prior for the
ellipticity of our mass model, i.e., we choose the SIE ellipticity to be lower than 0.1. This also implies that 
the model depends only weakly on the exact value of the ellipticity and PA. 
A flat prior is also given on the SIE position, using the astrometric error bars of Table~\ref{pos}. 
No priors are used for the other parameters. The best fit leads to $\chi^2=0.02$ and is obtained  for a SIE with 
$\sigma_{\rm lens}=215$ \kms, which converts into a total mass ${\rm M}_{\rm lens}=1.55 \times 10^{11}$~${\rm M}_\odot$.
The external shear is moderate, $\gamma=0.02$, with an position angle $\theta_\gamma=78^{\circ}$.   
A simple smooth model therefore accounts well for the astrometry of the quasar images, as also found
by previous studies. However, the predicted flux ratio between the quasar images (not included in the $\chi^2$
calculation) are not well reproduced at all.
 
We then add a perturber to the overall mass potential. We describe it as a Singular Isothermal Sphere (SIS) 
at the position of Object L. This adds extra free parameters to the models. In order to keep at least one degree of 
freedom in the fit, we successively fix the position of the lens galaxy to 25 different values within its astrometric uncertainty, and repeat the modelling. The remaining free parameters are the SIE and SIS velocity dispersions, the  SIS position, and the strength of the external shear, i.e., we have 5  free parameters vs.  6 observational constraints provided by 
the astrometry of the quasar images.
  
We then adopt flat priors on the SIS position using the astrometric error bars of Object L given in Table~\ref{pos}. 
The best fit has $\chi^2= 5.4$. The corresponding velocity dispersion for Object L is $\sigma_{L}= 0.3^{+4.0}_{-0.3}$ \kms (1-$\sigma$ error bar), which 
translates into a mass of ${\rm M}_{\rm L} < 2 \times 10^5$~${\rm M}_\odot$ (1-$\sigma$). The velocity dispersion of the main lens is unchanged
and the external shear is very little affected by the perturber, $\gamma= 0.018$.  
The mass we find for the perturber translates in a total mass-to-light ratio lower than unity in all bands. If we take the 3-$\sigma$ error bar on the velocity dispersion of the SIS as an upper limit 
the mass-to-light ratio of Object L is barely equal to unity. It is therefore unlikely that Object L is a satellite galaxy or 
even a globular cluster within the halo of the lens galaxy.

In addition, we note that introducing a perturber in the model at the position of Object L does not allow to 
reconcile  the astrometry of the quasar images and the observed flux ratio between quasar images A, D and C  as given  in Table~2 in \citet[][]{eigenbrod2006}. 
Adding more mass in Object L degrades the $\chi^2$ of the fit, i.e., it affects the astrometry, and still 
does not allow to reproduced  A/D flux ratio.


(ii) An alternative interpretation is that Object L is the 5$^{\rm th}$ central image of the quasar, like in \cite{Winn2004}.
Such a central image is 
produced when the mass profile of the lens galaxy is very shallow or truncated in its center. We describe this feature by introducing a core 
radius in our SIE model but we fail in predicting a central image at the position of Object L whatever is the size of the core 
radius. The hypothesis of a central image can therefore safely be ruled out.

    
 (iii) If Object L is the image of a source at the redshift of the quasar, we should see its counter image somewhere 
south to quasar image B and this counter image should  in principle  be brighter than Object L.  We do not see any evidence 
for a counter image in the PSF-subtracted AO data nor in the deconvolved HST data. Object L is most probably at a
redshift different from that of the quasar. If it is at a lower  redshift than the quasar,  we might see 
it in absorption in the quasar spectrum. None of the spectra available to us (i.e., SDSS\footnote{ Available at: http://cas.sdss.org/dr7/en/tools/explore/obj.asp?ra=0\\9:24:55.79\&dec=02:19:24.90} and VLT from Eigenbrod et al. 2006) show any trace 
of absorption lines at a different redshift than that of the lens galaxy. In addition, the very deep VLT spectrum of the lens galaxy includes
Object L in the slit and does not show any sign of objects at multiple redshifts. We therefore conclude 
that there is no evidence for Object L to be at a redshift different than that of the lens galaxy and that there
is no evidence for Object L to be lensed. 
      

(iv) The last simple explanation is that Object L is part of the lens galaxy but that it does not affect
much the lensing models. In the deconvolved ACS image, we detect a faint feature
almost aligned with Object L and the core of the lens galaxy. This feature, labelled ``\#4" in Fig.~\ref{figfitacs},
and Object L may reflect the presence of a bar in the lens galaxy. 
This is also supported by the fact that we need a disk component to model the light profile  
of the lens galaxy, both in the optical and in the near-IR.

\section{Conclusions}
\label{conc}

Motivated by the fact that lensed quasars with anomalous flux ratios  can be 
the signature of massive substructures in the lens galaxy, we have carried out deep VLT/AO 
imaging of \sdss, the quadruply imaged quasar with the most anomalous flux ratios known to date. 

With a total of 1.35~hours of exposure in the H-band and 1.4~hours in the Ks-band using the VLT NACO system
and the Laser Guide Star facility, we discover a luminous object  $\sim 0.3$\arcsec\ to the North 
of the lens galaxy. This object, that we call Object L, is also seen on archival HST/ACS images
in the F814W filter. 

We investigate the nature Object L using a smooth SIE model for the lens galaxy with an additional
SIS perturber at the same redshift and located the position of Object L. We show that such a model can not
explain the anomalous flux ratios in \sdss\ without also affecting the astrometry of the quasar images. 
The parameters found for Object L that allow a good fit of the astrometry all yield
total mass-to-light ratios smaller or barely equal to unity. 

We find evidence in the HST ACS/F814W images that Object L could be part of a  bar in the lensing
galaxy, oriented about along the N-S axis. This is consistent with our finding that the light profile of the 
lens galaxy can only be modeled with a 2-components distribution that includes a significant 
disk. 

We find no compact and luminous substructure that explains the flux ratio anomaly in \sdss. However the time scale of the anomaly is very large (7 years at least). Therefore it seems difficult to explain it only with micro-lensing.  A combination of milli-lensing and micro-lensing is the most likely interpretation for the observations. Observations of this system from UV to Mid-IR and,  such as in \citep{fadely2011}, would probably allow one to conclude about the origin of the flux anomaly in \sdss. 


\begin{acknowledgements} 
We acknowledge T. Broadhurst and S. Suyu for useful discussions. This work is partly supported by the 
Swiss National Science Foundation (SNSF). DS acknowledges partial support from the German Virtual Observatory and from the  Deutsche Forschungsgemeinschaft, reference SL172/1-1. DS thanks EPFL for financial support during his visit at the Laboratoire d'astrophysique where this work was finalized.
\end{acknowledgements}

\bibliographystyle{aa}
\bibliography{bibtex}

\end{document}